\UseRawInputEncoding  
\documentclass[aip,twocolumn,letterpaper]{revtex4}

\pdfoutput=1

\pdfoutput=1  

\usepackage{fullpage}
\usepackage{xcolor}

\usepackage{graphics}
\usepackage{epsfig}

\usepackage[margin=1.0in]{geometry}

\begin{document}
\title{Stellarators as a Fast Path to Fusion}
\author{Allen H Boozer}
\affiliation{Columbia University, New York, NY  10027\\ ahb17@columbia.edu}

\begin{abstract}

This paper is focused on three points:  (1) Overcoming obstacles to tokamak power plants may require a configuration modification as large as that of a stellarator.  (2) The demonstrated reliability of the computational design of stellarators should change fusion strategy.  (3) Deployment of carbon-free energy sources is mandated by the thirty-year doubling of carbon dioxide emissions.  Carbon-free energy options must be developed and fully deployed within a few doubling times. Unit size and cost of electricity are only relevant in comparison to alternative worldwide energy solutions. Intermittency, site specificity, waste management, and nuclear proliferation make fusion attractive as the basis for a carbon-free energy system compared to the alternatives.  Nonetheless, fusion is not an option for deployment until a power plant has successfully operated. A critical element in a minimal time and risk program is the use of computational design as opposed to just extrapolation.  Only the stellarator has an empirical demonstration of the reliable computational design through large changes in configuration properties and scale.  The computational design of stellarators should proceed while the inventions necessary for a more tokamak-like power plant are sought.  The cost of computational design is extremely small, but adequate time is required for the development of ideas that maximize attractiveness and minimize risk.  Rapid power-plant construction without many intermediate steps may seem risky, but the price is small compared to the cost of trillions of dollars  for each year's delay in addressing carbon-dioxide emissions. 

\end{abstract}

\date{\today} 
\maketitle


\begin{figure}
\centerline{ \includegraphics[width=3.0in]{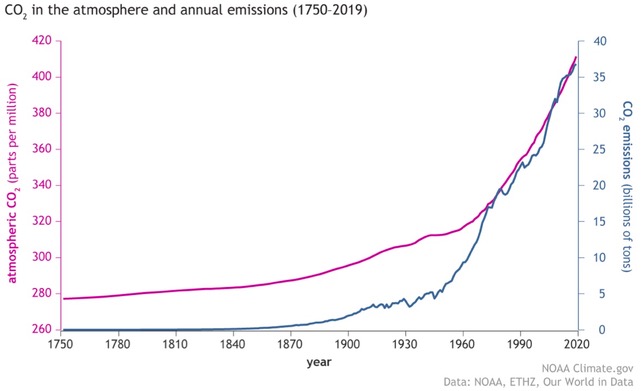} }
\caption{ The rate of CO$_2$ emissions is doubling approximately every thirty years and the enhancement in the atmospheric concentration above its pre-industrial level is doubling approximately every forty years.   This NOAA Climate.gov graph \cite{CO2} was adapted from the original by Dr. Howard Diamond (NOAA ARL). Atmospheric CO$_2$ data from NOAA and ETHZ. CO$_2$ emissions data from Our World in Data and the Global Carbon Project. }
\label{fig:CO2}
\end{figure}

\section{Introduction}

The situation in which fusion is being developed has undergone two fundamental changes.  (1) A general appreciation by the public that a doubling of the enhancement of the abundance of the carbon-dioxide in the atmosphere every forty years cannot be tolerated much longer.  (2) The demonstration that computational design of fusion plasmas confined by magnetic fields actually works \cite{Pedersen:W7-X} when the structure of the confining magnetic field is sufficiently dominated by external rather than by currents internal to the plasma. 

A sentiment is commonly stated without explanation \cite{FESAC:2021}: ``\emph{The tokamak approach for the plasma core is the most technically advanced and mature confinement concept.}"  Whatever may be meant by that statement, science implies the stellarator offers a far easier and more certain development path to fusion power on the shortest possible timescale.

Section \ref{sec:CO2} shows why carbon-dioxide issues mandate a rapid development of fusion.  Section \ref{sec:tokamak vs stellarator} explains obstacles that stand in the way of a tokamak power plant that can be circumvented using the stellarator.  It is unclear that smaller changes to the tokamak configuration could eliminate these obstacles.  Section \ref{comp design} considers the benefits of computational design, and Section \ref{discussion} is a summary.  The computational design of stellarators should proceed while the inventions necessary for a more tokamak-like power plant are sought.


\section{Carbon-dioxide mandate for fusion \label{sec:CO2}} 

The presently available options for addressing the problem of carbon-dioxide all have issues: intermittency, site specificity, waste management, and nuclear proliferation.  In principle, fusion energy is an extremely attractive solution, but a demonstration fusion power plant must be operated before that option can be judged.  The argument for minimizing the time and risk of building a fusion power plant is compelling.  The importance to world security far outweighs the financial cost.

Figure 1 shows that the rate of carbon-dioxide emissions is doubling approximately every thirty years, and the enhancement of the atmospheric concentration above its pre-industrial level has been doubling approximately every forty years.  World security requires something be done before too many doubling times.  A solution must be worldwide, or it is not a solution.  

Addressing the carbon-dioxide problem should be a two-step process.  First determine the options and then decide on which to deploy.  The cost of deployment is far greater that the research required to determine properties of an option.  Approximately a thousand power plants would be required for fusion to be a large fraction of the electricity generation capacity of the world, but in principle only one is required to determine the most important features.

The cost of all options, including doing nothing, is trillions of dollars per year.  Replacing the electricity generation capacity of the world and reaching the capacity required within thirty years has a capital cost of tens of trillions of dollars.  The cost of each year's emissions can be estimated from the \$50/ton estimate of the social cost of carbon used to set carbon taxes or from the less political number of \$200/ton, which is the estimated cost of direct CO$_2$ removal from the atmosphere \cite{NAS-CO2}.  Current emissions are $\approx 37\times10^9~$tons per year, which yields cost estimates of 1.8 to 7.4 trillions of dollars a year.  Ending the exponential increase in carbon-dioxide emissions and atmospheric concentration is a worldwide security issue and should be treated as such.

In 2018, the Intergovernmental Panel on Climate Change (IPCC) discussed the risks \cite{IPCC}  that  a temperature rise above $1.5^{\mbox{o}}$~C entails and argued for a fast implementation of solutions.  Wind and solar are generally taken to predominate energy production in a fast implementation, but there is a general recognition \cite{Burden} that the intermittency of wind and solar requires backup energy production that is cheap and can be quickly activated.  Gas turbines satisfy these requirements and are generally available at whatever scale is required.  Nuclear energy, fission or fusion, does not work well as an intermittent energy source.  Its high capital and low fuel costs works best for  providing an assured base load.  The intermittency of wind and solar coupled with their low cost per unit of energy make them ideal for energy intensive, interruptible operations in facilities with a low capital cost.  Such applications are not emphasized in reviews, but possibilities are the production of artificial fuels for zero-net-carbon energy systems and the direct removal of carbon dioxide from the atmosphere.   In any case, the low cost of demonstrating options relative to implementing them and the continual need to expand and replace energy production facilities imply the greater the perceived risk the more aggressive the research on potential options, such a fusion and direct removal of carbon dioxide, should be.

The importance of minimizing time and risk is so great that multiple fusion concepts could be pursued.  Nonetheless, one fusion concept is far better suited for the demonstration of fusion power with minimal time and risk: the stellarator.  The fundamental reason is that the magnetic configuration is dominated by the fields produced by external coils rather than by currents in the plasma.  This has three implications:  (1) Stellarator computational design has an unparalleled reliability \cite{W7-X:Nature,Pedersen:W7-X}. (2) Stellarator plasma maintenance and stability require singularly little active control.  (3) Stellarators have an enormous number of degrees of freedom due to their non-axisymmetry, which allows unique features and encourages invention.  

The combination of reliable computational design with an enormous number of degrees of freedom makes defining the problems of stellarators subtle; defined problems can generally be solved.  Stellarator design requires large-scale computers and an inventive spirit.  Stellarator research started seventy years ago in 1951.  In that era, the spirit of invention was strong but computers were weak.  Now that computers are strong, has the inventive spirit become too weak to achieve the types of things that could be achieved then?    It took approximately fifteen years to go from the splitting of uranium to fission powered submarines.  

How long would it take to develop fusion if it were considered of comparable  importance to security as the  development of nuclear fission was in 1939?  As was the case then, a rapid development of fusion power would require a carefully organized program  that is focused on societal needs and science.


\section{Tokamak issues---stellarator solutions \label{sec:tokamak vs stellarator}}


\subsection{Deuterium-tritium issues}

Deuterium-tritium fusion requires that a demonstration power plant (DEMO) meet a number of requirements.  The principal requirements are tritium self-sufficiency, tritium for initial start-up, and tritium safety.   A recent tokamak-focused review \cite{Tritium:2021} had a summarizing statement:
``\emph{A primary conclusion of this paper is that the physics and technology state of the art will not enable DEMO and future power plants to satisfy these principal requirements.}"  

Stellarators allow limitations cited in \cite{Tritium:2021} to be circumvented:

\subsubsection{Tritium breeding ratio (TBR)}

Page 7 of the DT fusion-cycle review \cite{Tritium:2021} notes: ``\emph{Physics requirements on the blanket in future fusion systems, such as presence of non-breeding materials for stabilizing shells, penetrations for heating, current drive, and other purposes, are not yet firmly established and can result in a substantial reduction in the achievable TBR.}" 

A fusing stellarator plasma can be maintained with zero current drive or heating power, which are required in tokamaks to  maintain not only the required plasma current but also the profiles of pressure and current.  The robust centering of stellarator plasmas within the plasma chamber and the absence of current driven kinks eliminates the need for stabilizing shells or coils.  The absence of disruptions and runaway electrons, Section \ref{disruptions-etc}, allows thinner walls.  

The adequate mitigation of disruptions and runaway electrons for ITER to achieve its mission is extremely challenging \cite{Plasma Steering,Eidietis:2021}, but they must be essentially eliminated for a power plant to be feasible \cite{Tritium:2021,DEMO-challenges,Eidietis:2021}.


\subsubsection{Burn fraction and fueling efficiency}

Page 13 of the DT fusion-cycle review \cite{Tritium:2021} notes: ``\emph{burn fraction and fueling efficiency represent dominant parameters toward realizing tritium self-sufficiency.}"  

As discussed in \cite{GA burn-frac} and Appendix \ref{Tritium burn}, the fraction of the tritium that is burned during one pass through a fusion plasma is small but can be controlled since tritium-burn fraction is inversely proportional to $f_t\equiv n_t/(n_t+n_d)$, the fraction of the ion density that is tritium.  The fusion power is proportional to $(1-f_t)f_t$ and hence the $nT\tau_E$ confinement requirement to maintain a burn is inversely proportional to $(1-f_t)f_t$, Appendix \ref{Tritium burn}.  The minimum $nT\tau_E$ confinement requirement is at $f_t=1/2$, but at $f_t=1/4$ the $nT\tau_E$ requirement is only 4/3 larger.  Section \ref{sec:Greenwald} explains why the tokamak Greenwald limit makes compensation more difficult in a tokamak than a stellarator. Power output is controllable by having fuel-injection pellets with differing tritium fractions.

The fueling efficiency might be enhanced in four ways in a stellarator.   (a) The shapes of stellarator plasmas offer places where pellet penetration is particularly easy.   (b) A stellarator power plant can in principle be designed so the transport is rapid in the inner part of the plasma with the required confinement for fusion given by an outer annulus \cite{CO2-Stell}.  The benefit is that the pellet needs to only penetrate the annulus, central aiming is not needed, and the danger of the pellet hitting the opposite wall is greatly reduced so faster pellets could be used.  (c) Stellarators generally do not have strong ELMs, which greatly degrade fueling efficiency. (d) Larger pellets have a smaller edge versus central ablation.  Page 30 of \cite{Tritium:2021} said: ``\emph{We expect shallow penetration of pellets in ITER and a DEMO plasma, even for a pellet with a content of 10\% of the plasma content, which is believed to be the limit of density perturbation that can be tolerated without adversely affecting the plasma.}''  Although the Greenwald limit \cite{Greenwald:2002} on the density in tokamaks does not constrain stellarators \cite{Greenwald:2002}, large oscillations in the density would cause large oscillations in the power output.  The optimal size for pellets requires study within the context of the overall power plant.


\subsubsection{Availability}

Page 10 of the DT fusion-cycle review \cite{Tritium:2021} notes:  ``\emph{Low availability factors could have tremendous consequences on tritium economy and self-sufficiency: during the reactor downtime...tritium production in blankets is interrupted whilst tritium is continuously lost by radioactive decay.}"   

A low availability can arise either from pulsed operation or from features that give a low reliability, availability, maintainability, or inspectability, which are called RAMI in the engineering literature.  The difficulty of achieving efficient maintenance of the net plasma current in a tokamak has led to a consideration of pulse operations, but this issue is not applicable to stellarators.  The absence of pulsing and the effects associated with the avoidance and mitigation of disruptions and runaway electrons are of great benefit to the RAMI issues of reliability and availability.  

Stellarators have the potential to be designed with coils that allow open access to the plasma chamber \cite{Yamaguchi:2019,CO2-Stell}, which is of central importance to the RAMI issues of maintainability and inspectability.


\subsection{Plasma steering, disruptions, and runaway electrons \label{disruptions-etc} }

Before a tokamak can have a successful power plant demonstration, the threat of plasma disruptions and runaway electrons must be eliminated \cite{DEMO-challenges,Eidietis:2021}.  This issue is often posed as an issue of plasma control or steering.  Unfortunately \cite{Plasma Steering}, ``\emph{Plasma steering sounds as safe as driving to work but will be shown to more closely resemble driving at high speed through a dense fog on an icy road.}"   Unlike stellarators, tokamaks require active control based on knowledge of the present state of the plasma but have few knobs with which to provide that control, and the diagnostics of the plasma state become far more limited in burning plasmas \cite{Diagnostics:2019}.  Eidietis recently wrote \cite{Eidietis:2021} ``\emph{The long-term mission of a reactor motivates investment in passive resilient design to survive disruptions in the absence of active intervention.}"

Stellarators have no issues that require a major invention, as elimination of disruptions and runaway electrons does in a tokamak \cite{CO2-Stell,Eidietis:2021}.  The structure of the magnetic field is dominated by the currents in the external coils, not in the plasma, which imparts a robust stability to stellarator configurations.  For example, position control of a tokamak plasma requires the vertical magnetic field evolve as the plasma equilibrium evolves---even feedback stabilization is generally required.   A change in the plasma state that is fast compared to the time scale for magnetic fields to penetrate through conducting structures can cause an irretrievable loss in position control.  In a stellarator, the externally produced magnetic fields ensure a plasma can never move far from its designed location in the plasma chamber.  In a power plant, the long time-delays of the plasma-control knobs place great importance on passive stability \cite{Eidietis:2021}.


\begin{table*}  
  \centering 
  \centerline{ \includegraphics[width=5.0in]{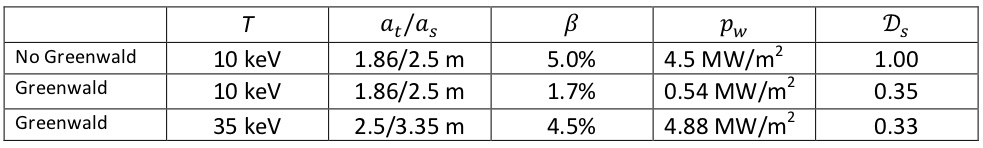} }
  \caption{The Greenwald limit, $n\leq n_G$, requires an increase in the plasma temperature $T$ to achieve an acceptable wall loading $p_w$ in a power plant and a decrease in the strength of the micro-turbulent transport, $\mathcal{D}_s$, is required to obtain ignition.   The assumed magnetic field is 6~T and the elongation $\kappa_e=1.8$.  For the cases in which the density equals the Greenwald density, the safety factor is $q_a=3$, the aspect ratio is $R/a_t=3$.  The first row gives typical stellarator power-plant parameters with no Greenwald limit.  The second row is identical to the first except the Greenwald limit is imposed.  The third row gives typical tokamak power-plant parameters with $n=n_G$ assuming a fixed $nT\tau_E$ requirement for a burning plasma.  The wall loading $p_w$ is similar in the first and third row, but the strength of the micro-turbulent transport $\mathcal{D}_s$ must three times smaller in the tokamak case despite having a plasma cross-sectional area, $\pi \kappa_ea_t^2=\pi a_s^2$, that is a factor of 1.8 times larger.   Actually between 14~keV and 35~keV, the $nT\tau_E$ requirement increases by a factor two, which makes the required $\mathcal{D}_s\approx0.17$ instead of 0.33 and the wall loading a factor of two lower, 2.44 instead of 4.88~MW/m$^2$.  The inclusion of synchrotron radiation would also significantly reduce the required $\mathcal{D}_s$ at 35~keV for a fusion burn.   }\label{Table}
\end{table*}

\subsection{Difficulties produced by the tokamak Greenwald density limit \label{sec:Greenwald} }

The avoidance of disruptions in tokamaks requires the line-averaged density in tokamaks be kept below \cite{Greenwald:2002} the Greenwald density, $n_G$.  The cause of the Greenwald limit is not universally agreed to, but it is related to the plasma current profile \cite{Cause density limit} and is not applicable in standard stellarators \cite{Greenwald:2002}.  

The Greenwald density limit greatly complicates the shutdown of tokamak plasmas \cite{Plasma Steering}, particularly burning plasmas, and as will be discussed, the design of divertors.  This section shows that the Greenwald limit forces power plants to operate at  a much higher plasma temperature \cite{Zohm:pp} than optimal and with a much smaller energy transport.   

When the plasma density $n$ in the plasma beta, $\beta=4\mu_0k_BnT/B^2$ , is replaced by $n/n_G$, the beta value with $n<n_G$ is very low unless the plasma temperature $T$ is extremely high,
\begin{eqnarray}
\beta&=& \frac{n}{n_G}\beta_G \mbox{   with   }
\beta_G\equiv c_G\frac{\kappa_e}{q_a}\frac{8k_BT}{BR},  \label{Greenwald beta}
\end{eqnarray} 
where $\beta_G$ is derived below.  The fusion power density scales approximately as beta squared,  Appendix  \ref{sec:wall loading}, so a high temperature is required to obtain a reasonable power density.  The energy confinement required for DT burning, $nT\tau_E$, is minimized at 14~keV.  Not only is the required $nT\tau_E$ a factor of two  larger when $T=35~$keV, but more importantly the empirical confinement in tokamaks and stellarators is gyro-Bohm like, which degrades as $T^{3/2}$.    Imposing the Greenwald limit forces a severalfold increase in the plasma confinement relative to gyro-Bohm.

The Greenwald density is defined by the plasma current $I_p$ and the half-width of the plasma in the midplane $a_t$, which in the tokamak  literature is called the plasma radius, $n_G\equiv c_G I_p/\pi a_t^2$.  In the International System of Units, the constant $c_G=10^{14}/$A$\cdot$m.  To obtain Equation (\ref{Greenwald beta}), the plasma current is expressed in terms of the safety factor at the plasma edge $q_a$, the plasma elongation $\kappa_e$, the magnetic field strength $B$ and the major radius $R$.  The result is the relation $n_G=c_G (\kappa_e/q_a)(2B/R\mu_0)$.  In the stellarator literature, the plasma radius $a_s$ is defined so $\pi a_s^2$ is the cross-sectional area; $a_s=\sqrt{\kappa_e}a_t$.

The empirical scaling law for energy transport in stellarators also works for tokamaks.  This is  shown in Figure 4 of \cite{W7-X:Nature}; the transport is essentially gyro-Bohm \cite{CO2-Stell}.   Consequently, the  strength of the transport can be quantified by multiplying gyro-Bohm by a dimensionless coefficient  $\mathcal{D}_s$.  The optimal level of transport for a power plant \cite{CO2-Stell} is $\mathcal{D}_s\sim1$. Nevertheless, as illustrated in Table \ref{Table}, the transport and $\mathcal{D}_s$ needs to be far smaller in a tokamak than a stellarator because of the Greenwald limit, even ignoring  the factor of two increase in $nT\tau_E$ due to the higher temperature. 

As shown in Table \ref{Table}, when $\beta_G$ is substituted for $\beta$ in Equation (\ref{wall loading}) of Appendix \ref{sec:wall loading} for the wall loading $p_w$,  a much higher that optimal plasma temperature $T$ is required for an adequate $p_w$, the fusion power per unit area on the walls.  When the same substitution is made into Equation  (\ref{PP conf}) of Appendix \ref{sec:PP conf}, a much smaller level of micro-turbulence $\mathcal{D}_s$ is required to be consistent with a power plant.


\section{Aspects of computational design \label{comp design}}

\subsection{Making fusion energy possible}

Until a fusion power plant has operated, the most important question is whether fusion energy is even feasible. 

The most important example of computational design removing a block to fusion feasibility is the 1988 demonstration by N\"uhrenberg and Zille \cite{QHS} that stellarators can have neoclassical transport that is consistent with a power plant.   Their work was based on a coordinate system developed by Boozer
\cite{Boozer:coord}, which generalized a simplified form for the guiding center particle drifts \cite{Boozer:drift H}.  The benefits of quasi-helical symmetry were seen in HSX the first stellarator built to study the N\"uhrenberg and Zille type configurations \cite{HSX-flow,HSX-transport}.

In non-axisymmetric plasmas, the guiding center drift velocity $v_{gc}$ can carry trapped particles a distance $v_{gc}\tau_c$ off the constant-pressure surfaces in a collision time $\tau_c$.  This gives the ripple-trapped neo-classical diffusion coefficient $D_{rt}\propto v_{gc}^2 \tau_c\propto T^{7/2}$, which can overwhelm the gyro-Bohm transport typically seen in both tokamaks and stellarators \cite{W7-X:Nature,CO2-Stell}, which scales as $T^{3/2}$. The ratio of the ripple-trapped to gyro-Bohm diffusion is proportional to the collisional mean-free-path divided the plasma size. This factor is greater than a thousand in a power plant, so ripple-trapped diffusion dominates unless the coefficient multiplying $v_{gc}^2 \tau_c$ is extremely small. Ripple-trapped diffusion occurs even in tokamaks because of the toroidal ripple, hence the name, and can be made small in stellarators---even zero on a magnetic surface which defines a confining annulus \cite{Garren:1991,CO2-Stell}.

The W7-X stellarator was made quasi-isodynamic rather than quasi-symmetric in order to minimize the parallel currents as well as the ripple-trapped diffusion \cite{Grieger:1992}.  The expected benefits of this optimization have been seen in the experiments \cite{Pedersen:W7-X,W7-X:Nature,7-X overview,7-X:High performance,7-X ambipolar}.  

W7-X shows \cite{Pedersen:W7-X} that transport due to particle drifts can be smaller than the gyro-Bohm transport and that this may be critical for the removal of impurities.  Typically ions are transported more rapidly than electrons by non-axisymmetry, and the resulting ambipolar electric field tends to pull impurities inward.  However, gyro-Bohm transport on both tokamaks and stellarators tends to clean the plasma of impurities.

 Optimization methods could be used to modify mico-turbulent transport in stellarators \cite{Xanthopoulos:2014}.  This is an area in which far faster computers or deeper physics understanding \cite{Hegna:2018}  could make a major contribution.

Axisymmetry eliminates ripple-trapped diffusion but comes at the price that the plasma must carry a current that is sufficiently strong to determine properties of the magnetic field.  This deprives the plasma of the passive stability and the maintenance properties that are the attraction of stellarators.  Unfortunately, theory and computation have not been able to provide solutions to these difficulties of axisymmetry.  The efficiency with which a plasma current can be driven has thermodynamic limits \cite{Boozer:current drive}.  The bootstrap current makes the maintenance of the current easier but also makes current-profile control far more subtle and the delicate stability situation in tokamaks even worse \cite{NC tearing}.  Major disruptions and runaway electrons, which must be eliminated before power plants become practical, have no clear solution \cite{Plasma Steering,Eidietis:2021}.  No solution is known to the Greenwald density limit \cite{Greenwald:2002} in tokamaks although there is no equivalent problem in stellarators \cite{Greenwald:2002}.

Divertor designs in stellarators are simplified by the location of the outermost confining magnetic surface being largely determined by currents in external coils rather than a balance between the hoop stress of the plasma current and pressure being balanced by a vertical field and by the absence of the Greenwald density limit.  The numerically determined island divertor \cite{Strumberger:divertor} on W7-X has worked as expected \cite{W7-X:divertor exp}  and has attractive features.  Numerical studies have also been carried out of a non-resonant divertor for stellarators, which unlike the island divertor, requires no particular rotational transform at the plasma edge \cite{Strumbeger:ergodic-div,Divertors-Boozer-Punjabi,HSX divertor:2017}.

A number of the issues discussed in Section \ref{sec:tokamak vs stellarator} are additional examples of how stellarator computer-design can address fusion feasibility issues.


\subsection{Shortening the time to fusion energy} 

Computational design can shorten the time to the operation of a power plant.  

Tokamaks have advanced by extrapolating from one generation of experiments to another.  The abstract of the original paper on the \emph{ITER Physics Basis} emphasized extrapolation \cite{ITER physics}.  Skepticism about the reliability of major design changes based on computation is well justified by the dependence of even basic features of a tokamak plasma on its non-linear, self-determined state.   The paper that introduced the scientific basis of W7-X emphasized computational optimization of designs \cite{Grieger:1992}.  The whole concept of design by computational optimization may seem ethereal to those accustomed to design by empirical extrapolation.  Nonetheless, computational design becomes reliable when the structure of the magnetic field is adequately dominated by the magnetic field produced by coils,  which can be calculated using the Biot-Savart Law.

Extrapolation has a number of disadvantages in comparison to computational design \cite{CO2-Stell}:
\begin{enumerate}

\item Experiments build in conservatism---even apparently minor changes in design are not possible and therefore remain unstudied.  Major changes are risky even when going from one generation of experiments to another

\item  Experiments are built and operated over long periods of time---often a number of decades.  A fast-paced program is inconsistent with the need for many generations of experiments.

\item  The cost of computational design is many orders of magnitude smaller than building a major experiment, as well as having a much faster time scale. 

\item  Extrapolations are dangerous when changing physics regimes.  Examples are (i) plasma control in ignited versus non-ignited plasmas and (ii) the formation of a current of relativistic electrons during a disruption. 
 \end{enumerate}


\subsection{Making fusion energy more attractive}
Contrary to common opinion, once an appropriate stellarator configuration is found, it need not be more sensitive to magnetic field errors than tokamaks. Stellarators do not suffer from the large amplification of the error fields that occurs in tokamaks due to weakly-stable current-driven kinks \cite{Park-error}.  Nonetheless, magnetic field errors can cause problems.  The issue can be addressed by careful stellarator construction \cite{Pedersen:W7-X errors} as in W7-X, but this is expensive.  Error-field control coils can be used as they are in tokamaks.  Condition-number issues of the matrices that relate the control-coil currents to the error-field magnetic distributions limit the number of error fields that can be controlled and careful design \cite{Boozer:design} should be used to minimize the cost of construction.  The same design techniques that give optimal error-field control also allow stellarators to be designed for optimal flexibility in physics studies.  The basic principle is that any external magnetic field to which the physics is sensitive should be controllable by adjusting the currents in the external coils.

Curl-free magnetic fields can be designed that provide excellent physics properties.  Nonetheless, a practical stellarator must have a plasma pressure $p$ or beta, $\beta\equiv2\mu_0 p/B^2$.  The DT fusion power density is approximately proportional to pressure squared \cite{Wesson:2004}, Appendix \ref{sec:wall loading}; the required plasma pressure is determined by the required fusion power density for economic fusion power.  The higher the beta, the more efficient is the utilization of the magnetic field.  Stellarators can be designed and have beta limits above the 5\% often said to be required for a reactor, but the actual limit seems to be soft \cite{Xu-tok-stell} and may in practice be far higher.  The computational design of stellarators is more reliable when the plasma beta is below all theoretical limits, but this may unnecessarily increase the cost of stellarator power plants.  In principle, the required magnetic field can be studied on a demonstration power plant by reducing the magnetic field produced by the coils below the design value.

As discussed, the coil system on a stellarator can be apparently designed for open access.  How this should be optimized to simplify maintenance and repair remain to be studied.


\section{Discussion \label{discussion}}

The ITER project was set in motion more than a third of a century ago, in 1985 by Ronald Reagan and Mikhail Gorbachev. At that time, neither the solution \cite{QHS} to the problem of ripple transport in stellarators (1988) nor the limits \cite{Boozer:current drive} on current drive efficiency (1988) and plasma density in tokamaks (1988) were known. The tokamak community was not familiar (1997) with the dangers of runaway electrons \cite{R-P:runaway}, nor was the urgency of ending the doubling of carbon-dioxide generally appreciated. Science and society have evolved during the last third of a century.  The fusion program must evolve as well.

The costs associated with carbon-dioxide emissions, which are trillions of dollars a year, necessitate a quick determination of options for solutions.  For minimization of time and risk, the robustness of the stability and maintenance of the plasma are paramount.  An invention is required for a tokamak to have sufficient passive stability for a power plant.  Is the stellarator the required invention that makes the least change in the fundamental concept?  The computational design of stellarators should proceed even while  inventions necessary for a more tokamak-like power plant are being sought.

 \appendix
 
 \section{DT power and plasma confinement}
 
Convenient units are 10~keV for temperature, $10^{20}/$m$^3$ for density, Tesla for magnetic fields, megaampers for current, megajoules for energy, seconds for time, and meters for distances.  In these units, the Boltzmann constant $k_B=0.1602$, the permeability of free space $\mu_0 = 0.4\pi$, and the Greenwald constant $c_G=1$.   
 
 \subsection{Tritium-burn fraction \label{Tritium burn}}

The rate at which tritium is burned is $dn_t/dt=-n_tn_d\langle\sigma v\rangle_{dt}$.  The burn-time of the tritium $\tau_{tb}$ is defined so $n_t\propto\exp(-t/\tau_{tb})$, so $\tau_{tb} \equiv 1/(n_d\langle\sigma v\rangle_{dt})$.

 The fraction of the tritium burned before it is swept out of the plasma on the  tritium-particle confinement time $\tau_p$ is $f_{tb} \equiv 1-  \exp(-\tau_t/\tau_{tb})$.   Since the tritium-burn fraction $f_{tb}$ is much less than unity, $f_{tb} \approx   \tau_t/\tau_{tb}$.  This relation will be treated as if it were exact.
 
 The power density of the plasma heating by the alpha particles $p_\alpha$ and the energy confinement time $\tau_E$ required to maintain a burning plasma are  $p_\alpha=\mathcal{E}_{\alpha} n_tn_d\langle\sigma v\rangle_{dt}$ and $\tau_E=3k_BnT/p_\alpha$, where $\mathcal{E}_{\alpha}=3.5~$Mev is the energy release in an alpha particle per reaction, $n=n_d+n_t$ and $T$ is the plasma temperature.  The $3nk_BT$ is the thermal energy of the electrons plus the ions. 
 
The fraction of the tritium burned in one pass through the plasma is then
\begin{equation}
f_{tb} = \frac{1}{f_t}   \frac{\tau_t}{\tau_E}   \frac{3  k_BT }{ \mathcal{E}_{\alpha}}.
\end{equation}
The smallness of the tritium-burn fraction is primarily due to $3k_BT/\mathcal{E}_{\alpha}$ being of order one percent.

The tritium-burn fraction can be increased by making $\tau_t/\tau_E$ large or the tritium fraction $f_t$ small.  A reduction in $f_t$  requires making the energy-confinement time $\tau_E$ longer to satisfy the condition for a fusion burn,
\begin{eqnarray}
 nT\tau_E &\equiv& \frac{3}{f_t(1-f_t)} \frac{k_BT^2}{\mathcal{E}_{\alpha}\langle\sigma v\rangle_{dt}}.
 \end{eqnarray}
 Although a large $\tau_t/\tau_E$ is beneficial, long confinement time for the fusion-produced $\alpha$ particles is not.  Their density can be shown to be \cite{burn-fraction}
 \begin{eqnarray}
\frac{n_\alpha}{n}&=& \frac{\tau_\alpha}{\tau_E} \frac{3k_BT}{\mathcal{E}_{\alpha}}.
\end{eqnarray}
Unless $n_\alpha/n$ is small, the $\alpha$ particles unacceptably enhance the plasma pressure and the bremsstrahlung power loss. 

 An energy confinement time  $\tau_E$ that is too short can be compensated by increasing the magnetic field and plasma size.  However, a ratio $\tau_t/\tau_E$ that is too short or a $\tau_\alpha/\tau_E$ ratio that is too long are presumably more difficult to modify.  It is important that the physics that determines these ratios be studied both theoretically and empirically.

 
 \subsection{Wall loading \label{sec:wall loading}}

Wesson  \cite{Wesson:2004} noted that  $T^2/\langle\sigma v\rangle_{dt}$ is at a minimum at $T=14~$keV and  only weakly dependent on $T$ in the temperature range of primary interest for fusion.  His result is that power per unit volume released in neutrons and alpha particles in a plasma with equal deuterium and tritium densities is $p_{DT}=5c_{DT}n^2T^2$ with $c_{DT}=0.154$.  The economics of fusion is dependent on obtaining an adequate power per unit area on the walls, $p_w = (p_{DT}\pi \kappa_e a_t^2)/((1+\kappa_e)\pi a_t)$.  The plasma beta is $\beta \equiv 2\mu_0p/B^2$ so $nT=\beta B^2/4\mu_0k_B$, so the nuclear power per unit area on the walls is
\begin{eqnarray}
p_w &=&\frac{5}{16} \frac{\kappa_e}{1+\kappa_e}  \frac{c_{DT}}{(\mu_0 k_B)^2} \beta^2 B^4 a_t. \label{wall loading}
\end{eqnarray}

\subsection{Required plasma confinement \label{sec:PP conf}}

A power plant requires the alpha particle heating balance the energy losses, which means $c_{DT}(nT)^2=3k_BnT/\tau_{E}$, where $\tau_E$ is the confinement time for energy.  Equivalently, $nT\tau_E=3k_B/c_{DT}\approx3.16$.

As discussed in the Appendix to \cite{CO2-Stell} and illustrated in Figure 4 of \cite{W7-X:Nature}, the empirical energy confinement in stellarators and tokamaks is similar and gyro-Bohm like.  The gyro-Bohm diffusion coefficient, $D_{gB} \equiv \rho_i^2 C_s/a_t$, where $C_s\equiv\sqrt{T/m_i}$ and the ion gyroradius is $\rho_i\equiv C_s/\omega_{ci}$, so $D_{gB}=c_{gb} T^{3/2}/a_tB^2$ with $c_{gB}\approx162$.  The dimensionless strength of the micro-turbulent transport is defined by $\mathcal{D}_s$, with $nT\tau_E=nTa_t^2/\mathcal{D}_s D_{gB}$, or 
\begin{eqnarray}
\mathcal{D}_s &\equiv& \frac{n}{(nT\tau_E)} \frac{B^2a^3}{c_{gB} T^{1/2}}, \mbox{    and   } \label{conf}\\
&=&  \frac{c_{DT}}{12\mu_0k_B^2c_{gB}} \frac{\beta B^4a_t^3}{ T^{3/2}} \label{PP conf}
\end{eqnarray}
using the required $nT\tau_E$ for a burning plasma.  An empirical $\mathcal{D}_s$ can be obtained from experiments using Equation (\ref{conf}).

\vspace{0.2in}

\section*{Acknowledgements}

This material is based upon work supported by the U.S. Department of Energy, Office of Science, Office of Fusion Energy Sciences under Award DE-FG02-95ER54333 and by grant 601958 within the Simons Foundation collaboration ``\emph{Hidden Symmetries and Fusion Energy}."


\end{document}